\documentstyle[12pt]{article}
\title{\nopagebreak
\begin{flushright}
\tenrm UCTP112.98
\end{flushright}\vskip0.3in
\nopagebreak
\large \bf Black Holes and Super Black Holes \\ as \\
Chern Simons Theories in 2+1 Dimensions}
\author{Sharmanthie Fernando\thanks{email address:
fernando@physung.phy.uc.edu} and Freydoon Mansouri\thanks{email
address: Mansouri@uc.edu} \\
\it \small \it Physics Department, University of Cincinnati,
Cincinnati, OH 45221}
\date{}

\begin{document}
\maketitle

\begin{abstract}

We study 
anti-de Sitter black holes in 2+1 dimensions in terms of Chern
Simons gauge theory of the anti-de Sitter group coupled to a
source. Taking the source to be an anti-de Sitter state specified
by its Casimir invariants, we show how all the relevant features
of the black hole are accounted for. Enlarging the gauge symmetry
to super AdS group, we obtain a supermultiplet of AdS black
holes. We give explicit expressions for masses and the angular
momenta of the members of the multiplet.
\end{abstract}

\section{introduction}
The AdS black hole in 2+1 dimensions is a solution of free
Einstein's equations with a negative cosmological
constant~\cite{rone}. It is well known that the free Einstein
theory in 2+1 dimensions with or without a cosmological constant
can be formulated as a free Chern Simons
theory~\cite{rtwo,rthree}. In a free Chern Simons theory, the
field strength vanishes identically, so that there are no local
degrees of freedom. It is then somewhat surprising that a theory
with no local degrees of freedom has a black hole solution which
must have non-trivial degrees of freedom to account for its
entropy! Presumably, M theory, once it is fully constructed, will
provide us with the correct answer. In the meantime, one of the
suggestions to resolve this issue is to modify the Chern Simons
action
with a WZW term~\cite{rfour,rfive}. Another suggestion is to make
use of the AdS/CFT correspondence~\cite{rsix}. More recently, it
has been suggested that the answer lies in the coupling of the
Chern Simons theory to a source~\cite{rseven}. In this approach,
the free Chern Simons theory is not only taken to be locally
trivial. It is
also taken to be globally trivial in accord with Mach's
principle. Non-trivial topologies then arise as a result of
coupling to sources.
This is one of the issues which we will address in this work.

One of the notable advantages of this approach is
that it allows us to express the asymptotic observables of the
theory in terms of the properties of the sources. To implement
this idea, we must identify a localized source (particle) with an
irreducible representation of the gauge symmetry
group~\cite{reight}. For the present problem, this will amount to
relating the asymptotic observables of the BTZ black hole to the
Casimir
invariants of an AdS state coupled to the Chern Simons action. We
will show that
space-time will naturally emerge from such a gauge theory
and will have all the ingredients
necessary for the AdS black hole~\cite{rone,rnine}. These
include, in
particular, the discrete subgroup underlying the identifications.

A second issue which we will address in this work relates to the
manner in which black hole solutions fit into supersymmetric
schemes. 
A conventional method of searching for signs of supersymmetry in
black hole solutions is to look for Killing spinors. Many
works along these lines already exist in the
literature. We cite a representative few
here~\cite{rten}-~\cite{reighteen}, from which more references
can be traced. 
One way to see whether a given black hole solution admits Killing
spinors is identify it with the bosonic part
of an appropriate
supergravity theory~\cite{rten}-~\cite{reighteen}. Then, by
requiring
that the fermions in the theory as well
as their variations vanish, one arrives at Killing spinor
equation(s). The asymptotic supersymmetries depend on the number
of non-trivial solutions of these equations consistent with the
black hole topology. For example, in asymptotically flat
space-times, a
typical supermultiplet consists of a black hole and a number of
ordinary particles all with the same mass. In contrast to the
familiar situation in particle physics, where we have
Supermultiplets consisting of particles only, in this approach
there is no systematic way of looking for
supermultiplets consisting of black holes only. We will show that
in $2+1$
dimensions it is possible to construct a theory which permits
macroscopic solutions consisting of all AdS black hole
supermultiplets~\cite{rnineteen}. It
involves the Chern Simons gauge theory of the (1,1) super AdS
group coupled to a super AdS state
(source). As we shall see below, to be able to accommodate the
structure of the solution which emerges from such a theory it
becomes necessary to broaden the standard notions of classical
geometries to include some quantum mechanical elements.

\section{Anti-de Sitter space and algebra}
The anti-de Sitter space in 2+1 dimensions can be viewed  as a
subspace of a
flat 4-dimensional space with the line element
\begin{equation}
ds^2 = dX_AdX^A = dX_0^2 - dX_1^2 -dX_2^2 + dX_3^2 \end{equation}
It is determined by the constraint
\begin{equation}
(X_0)^2 - (X_1)^2 - (X_2)^2 + (X_3)^2 = l^{2}\end{equation}
where $l$ is a real constant . The set of transformations which
leave the line element invariant
form the anti-de Sitter group $SO(2,2)$. It is locally isomorphic
to $SL(2,R) \times SL(2,R)$. From
here on by
anti-de Sitter group we shall mean its universal covering group.

With $a= 0,1,2,$ we can write the AdS algebra in two
more convenient forms:
\begin{eqnarray}
\epsilon^{abc}J_c  = \epsilon^{abc} 
( J^+_c +J^-_c ) \nonumber \\
l\Pi^a = ( J^{+a} - J^{-a} ) \end{eqnarray}
where
\begin{equation} 
\epsilon^{012} = 1 ; \hspace{1.cm} \eta^{ab} = (1, -1,
-1)\end{equation}
Then, the commutation relations in, say, $J_a^{\pm}$ basis take
the form~\cite{rseven} 
\begin{equation}
\left[J_a^{\pm}, J_b^+\right] = -i\epsilon_{ab}^c
J^{\pm}_{c};\;\;\;\;\;
\left[J_a^+, J_b^-\right] = 0 \end{equation}
The Casimir operators in this basis have the form
\begin{equation}
j_{\pm}^2 = \eta^{ab} J^{\pm}_a J^{\pm}_b
\end{equation}
Alternatively, we can use a combination of these with eigenvalues
corresponding to the parameters of the BTZ solution:
\begin{eqnarray}
M = l^2 (\Pi^a\Pi_a + l^{-2}J^aJ_a) = 2(j_{+}^2 + j_{-}^2)
\nonumber \\
J/l = 2l\Pi_aJ^a = 2( j_+^2 - j_-^2)\;\;\;\;\;
\end{eqnarray}
Unless otherwise stated, we will use the same symbols for
operators and their eigenvalues.

An irreducible representation of AdS group can be labeled by the
eigenvalues
of either the pair $(M,J)$ or the pair $(j_+,j_-)$. For
application to black holes, it is also possible to label these
representations by eigenvalues which are proportional to the
horizon radii $r_{\pm}$ of the AdS black hole~\cite{rone,rseven}.

\section{Connection and the Chern Simons action}

To write down the Chern Simons action, we begin by writing the
connection in $SL(2,R) \times SL(2,R)$
basis.
\begin{equation}
A_{\mu} = \omega ^{AB}_{\mu} M^{AB} = \omega ^a_{\mu} J_a +
e^a_{\mu} \Pi_a
= A^{+a}_{\mu} J_a^+ + A_{\mu}^{-a} J_a^- \end{equation}
where 
\begin{equation}
A^{\pm a}_{\mu} = \omega ^a_{\mu} \pm l^{-1} e^a_{\mu}
\end{equation}
Eq. 9 should be viewed as definitions of $e$ and
$\omega$ in terms of the two $SL(2,R)$ connections. The covariant
derivative can be written as
\begin{equation}
D_{\mu} = \partial_{\mu} - iA_{\mu}= \partial_{\mu}
-iA^{+a}_{\mu} J_a^+ -i A_{\mu}^{-a} J_a^- \end{equation}
Then the components of the field strength are given by
\begin{equation}
[D_{\mu}, D_{\nu}] = -iF^{+a}_{\mu \nu} J^+_a - iF^{-a}_{\mu
\nu} J^-_a 
 = -iF^{+}_{\mu \nu}[A^+] -  iF^{-}_{\mu \nu}[A^-] \end{equation}

For  a simple or a semi-simple group, the Chern Simons action has
the form
\begin{equation}
I_{cs} = \frac{1}{4\pi}Tr \int_M A \wedge \left( dA + \frac{2}{3}
A \wedge A\right) \end{equation}
where Tr stands for trace and
\begin{equation}
A = A_{\mu} dX^{\mu} =  A^+ + A^- \end{equation} 
We require the 2+1 dimensional manifold M to have the topology 
$R\times\Sigma$, with $\Sigma$ a two-
manifold.
So, The Chern Simons action with $SL(2,R)\times SL(2,R)$ gauge
group will take the form 
\begin{equation}
I_{cs} = \frac{1}{4\pi}Tr \int_M \left[\frac{1}{a_{+}}A^+ \wedge
\left( dA^+ +
\frac{2}{3}
A^+ \wedge A^+ \right) + \frac{1}{a_{-}}A^- \wedge \left( dA^- +
\frac{2}{3}
A^- \wedge A^- \right) \right] \end{equation}
Here the quantities $a_{\pm}$ are, in general, arbitrary
coefficients, reflecting the semisimplicity of the gauge group.
Up to an overall normalization, only their ratio
is significant. As explained elsewhere~\cite{rseven}, in the
presence of a source (or of sources), any
\'a priori choice of the coefficients $a_{\pm}$ reduces the class
of allowed holonomies, so that even the classical theory coupled
to sources will be
affected by such a choice. For this reason, we will keep the
coefficients $a_{\pm}$ as free parameters in the sequel, so that
we can generate the correct holonomies for solutions both outside
and inside the horizon. 

Under infinitesimal gauge transformations
\begin{equation}
u_{\pm} = \theta^{\pm\;a} J^{\pm}_{a} \end{equation}
the gauge fields transform as
\begin{equation}
\delta A_{\mu} = - \partial_{\mu} u - i[ A_\mu,u] \end{equation}
More specifically,
\begin{equation}
\delta A^{\pm\;a}_{\mu} = -\partial_{\mu}\theta^{\pm\;a} -
\epsilon^{a}_{\;bc}A^{\pm\;b}\theta^{\pm\;c}\end{equation}

As we have stated, the manifold $M$ has the topology $R \times
\Sigma$ with R representing $x^{0}$. Then subject to the
constraints
\begin{equation}
F^{\pm}_a[A^{\pm}] =\frac {1}{2}  \eta_{ab} \epsilon^{ij}
(\partial_i
A_j^{\pm\;b} - \partial_j A_i^{\pm\;b} + \epsilon^b_{\;cd}
A_i^{\pm\;c}
A_j^{\pm\;d}) = 0 \end{equation}
the Chern Simons action for $SO(2,2)$ will take the form
\begin{eqnarray}
2\pi I_{cs} = \frac{1}{a_+} \int_R dx^{0}  \int_{\Sigma}
d^2x\left(- 
\epsilon^{ij}\eta_{ab} A^{+a}_i \partial_0 A^{+b}_j +   A^{+a}_ 0
F^+_a \right)\nonumber \\
+ \frac{1}{a_-} \int_R dx^{0}  \int_{\Sigma} d^2x\left(- 
\epsilon^{ij}\eta_{ab} A^{-a}_i \partial_0 A^{-b}_j +  A^{-a}_0
F_a^- \right)\end{eqnarray}
where $i,j = 1,2$.

\section{$SO(2,2)$ States and Interaction with sources}

Following the approach which has been successful in coupling
sources to Poincar\'e Chern Simons
theory~\cite{reight}, we take a source for the present problem to
be an irreducible representation
of anti-de-Sitter group characterized by Casimir invariants $M$
and $J$ (or $r_+$ and $r_-$ ). 
Within the representation, the states are further specified by
the phase space
variables of the source $\Pi^A$ and $q^A$, $A= 0,1,2,3$, subject
to anti-de Sitter constraint given by Eq. 2. To allow for the
possibility
of quantizing the Chern Simons theory consistently, we must
require that our sources be represented by unitary
representations of the AdS group. The choice of relevant
representations from among these were discussed in reference [7].
Here we note that since the AdS group in 2+1 dimensions can be
represented in the $SL(2,R) \times SL(2,R)$ form, the unitary
representations of $SO(2,2)$ can be constructed from those of
$SL(2,R)$. The latter group has four series of unitary
representations all of which are infinite dimensional.
Of these, the appropriate representations turn out to be the
discrete series bounded from below~\cite{rseven}. The states in
an
irreducible representation of $SL(2,R)$ are specified by the
eigenvalues of its Casimir operator and one of the generators. In
reference~\cite{rseven}, the compact generator was chosen to
label the
states. This corresponds to inducing representations of $SL(2,R)$
via its maximal compact subgroup $SO(2)$. Thus, suppressing the
superscripts $\pm$ which distinguish our two $SL(2,R)$'s, we can
write
$$j^2 |F, m> = F (F - 1) |F, m>$$
$$J_0 |F, m> = (F + m) |F, m>$$
In these expressions
\begin{equation}
F = real \;\; number \geq 0; \;\;\;\;\; m
= 0, 1, 2,...
\end{equation}
So, for this series,
the eigenvalues of the Casimir
invariants of $SL(2,R) \times SL(2,R)$
can be written as, 
\begin{equation}
j^2_{\pm} = F^2_{\pm} - F_{\pm}
\end{equation}
It follows that the infinite set of states can, in a somewhat
redundant notation, be
specified as
\begin{equation}
 |j_{\pm}^2, F_{\pm}+ m_{\pm}>; \;\;\;\;\; m_{\pm} = 0, 1, 2, ...
\end{equation}
Clearly, the integers $m_{\pm}$ are not necessarily equal.
Using these states, we can construct the discrete series of the
unitary
representations of $SO(2,2)$. A typical state
will have the following labels:
\begin{equation}
| M, J > = | j_+^2, j_-^2, F_+ + m_+, F_- + m_- >
\end{equation}
To be able to identify the labels $M$ and $J$ with the
corresponding labels in the AdS black hole, we must require that
$F_{\pm} \geq 1$~\cite{rseven}. It would then follow that
$|J/l| \leq
|M|$, as required for having a black hole solution.

The main advantage of diagonalizing a compact generator is that
the maximal compact subgroup $SO(2) \times SO(2)$ of $SO(2,2)$
allows a parametrization of the AdS space in terms of circular
functions so that the corresponding angular variable is periodic
to begin with and will remain so in the black hole solution. As
a result~\cite{rseven}, the periodicity necessary to obtain the
discrete identification group need not be imposed as an
additional requirement as was done by BTZ~\cite{rone}. However,
to obtain the BTZ solution from this starting point, it becomes
necessary~\cite{rone} to perform a ``Wick rotation'' in the space
of Casimir invariants to recover the hyperbolicity of the angular
variable. An alternative to this procedure is to construct the
SL(2,R) representations by diagonalizing a non-compact generator.
In that case, the irreducible representations of $SO(2,2)$ can be
constructed via induction from the subgroup $SO(1,1) \times
SO(1,1)$. One way to do this is to make the formal replacement
$J_0 \rightarrow iJ_2$ to go from the formalism in which the
compact generator $J_0$ is diagonal to one in which the non-
compact generator $J_2$ is diagonal . Obviously, the structure
and the
classification of the unitary representations will remain the
same. The only difference is that in the latter case the angular
variables to be used to
parametrize the AdS space are hyperbolic to begin with. Then, as
we shall see below, the
$2\pi$ periodicity of the hyperbolic angular variable will arise
from the requirement of compatibility with the topological
properties of the Chern Simons theory coupled to a source.

With these preliminaries, we can couple a source to the AdS Chern
Simons theory in the following way:
$$I_{s}  = \int_C d \tau\left[\Pi_A \partial_{\tau}q^A -
(A^{+a}J^+_a + A^{-a}J^-_a)+ \lambda\left( q^Aq_A - l^2
\right)\right]$$ 
\begin{equation}
+\int_C d\tau\left[ \lambda_+ \left( J^{+a} J^{+}_a -
l^2j_+^2\right)+
\lambda_-
\left(J^{-a} J_a^- - l^2j_-^2\right)\right]
\end{equation}
In this expression, $C$ is a path in $M$, $\tau$ is a  parameter
along $C$, and $J^{\pm}_a$ play the role of c-number
generalized
 angular momenta which transform in the same way as the
corresponding generators which label the source. The quantities
$\lambda$ and $\lambda_{\pm}$ are Lagrange multipliers.
The first constraint in this action ensures that $q^A(\tau)$
satisfy the AdS
constraint. As explained in previous
occasions~\cite{rseven,reight,rtwenty}, it is
not the manifold $M$ over which
the gauge theory is
defined but the space of $q_A$'s which give rise to the classical
space-time. The last 
two constraints identify the source being coupled to the Chern
Simons theory as an
anti-de Sitter state with invariants $j_+$ and $j_-$. These
constraints are crucial in relating
the invariants of the source to the asymptotic observables of the
coupled theory via Wilson loops.

The total action for the theory is given by:
\begin{equation}
I = I_{cs} + I_{s} \end{equation}
It is easy to check that in this theory
the components of the field strength still vanish
everywhere except at the location of the sources.
So, the analog of Eq. ? become
\begin{equation}
\epsilon^{ij} F^{\pm\;a}_{ij} =2\pi a_{\pm} J^{\pm\;a}
\delta^2(\vec{x},\vec{x_0})
\end{equation}
In particular, fixing the gauge so that $SO(2,2)$ symmetry
reduces to $SO(1,1) \times SO(1,1)$, we get
\begin{equation}
\epsilon^{ij} F^{\pm\;2}_{ij} = 2\pi a_{\pm} F_{\pm}
\delta^2(\vec{x},\vec{x^0})
\end{equation}
where $F_{\pm}$ are the invariant labels of the state as in Eq.
23, but now they are associated with the non-compact generator
$J_2$.
All other components of the field strength vanish. We thus
see that because of the 
constraints appearing in the action given by Eq. 24, the strength
of the sources become related 
to their Casimir invariants. These invariants,
in turn, determine the asymptotic observables of the theory.
Since such 
observables must be gauge invariant, they are expressible in
terms of Wilson loops,
and a Wilson loop about our source can only depend on, e.g.,
$j_+$ and $j_-$.

From the data on the manifold $M$ given above, one can
determine the properties of the emerging space-time by solving
Eqs 27. The only
non-vanishing components of the gauge potential are given
by~\cite{rseven}
\begin{equation}
A^{2 \pm}_{\theta} = 2a_{\pm} F_{\pm}
\end{equation}
where $\theta$ is an angular variable. Although these are
components of a connection which is pure gauge, they give rise to
non-trivial holonomies around the source. More explicitly, we
have
\begin{equation}
\omega[A^+] = exp^{\pi(r_-+r_+) J_2^+}
\end{equation}

and
\begin{equation}
\omega[A^-] = exp^{\pi(r_--r_+) J_2^-}
\end{equation}
Since we are diagonalizing
$J^{\pm}_2$ operators, their matrix representation is given by
\begin{eqnarray}
J^{\pm}_2(\alpha) = \frac{1}{2} \left(\begin{array}{cc} 1  &  0
\\  0 & -1 )
\end{array} \right) 
\end{eqnarray}
Then, the above holonomies will take the form
\begin{eqnarray}
\omega[A^+] =  \left(\begin{array}{cc} exp^{\pi(r_-+r_+)}  &  0
\\  0 & exp^{-\Pi(r_-+r_+) }
\end{array} \right) 
\end{eqnarray}  
\begin{eqnarray}
\omega[A^-] = \left(\begin{array}{cc} exp^{\pi(r_+-r_-)}  &  0 \\ 
0 & exp^{-\Pi(r_+ -r_-) }
\end{array} \right) 
\end{eqnarray}  
It was shown in
reference [7] how these holonomies lead to a discrete
identification subgroup of $SO(2,2)$, indicating that the
manifold $M_q$ of the $0+1$ dimensional fields $q^A$ has all the
relevant features of the macroscopic AdS black hole solution. As
we shall see below, the same holonomies, suitably interpreted,
will play a crucial role in establishing the space-time structure
of the supersymmetric theory discussed below.

\section{The black hole space-time}
To see how the space-time structure emerges from our anti-de
Sitter gauge theory, we follow an approach which led to
the emergence of space-time from
Poincar\'e~\cite{reight} and super Poincar\'e~\cite{rtwenty}
Chern Simons gauge
theories. We have emphasized that the manifold $M$ is not to be
identified with space-time. But the information encoded in $M$
and discussed in the previous section is sufficient to fix the
properties of
the emerging space-time. To this end, let us consider a manifold
$\hat{M}_q$ satisfying the AdS constraint
\begin{equation}
\hat{q}_0^2 - \hat{q}_{1}^2 - \hat{q}_2^2 + \hat{q}_3 ^2 = l^2 =
-\Lambda^{-1}
\end{equation}
where $\Lambda$ = cosmological constant. In fact, our $SL(2,R)
\times SL(2,R)$ formulation allows us to take $\hat{M}_q$
to be the universal covering
space of the AdS space. As we shall see, the emerging space-time
is the quotient of $\hat{M}_q$ by the discrete subgroup
$\Gamma$
discussed in the previous section. Moreover, the source coupled
to the Chern
Simons action is an AdS state characterized
by the Casimir invariants $(M,J)$ or, equivalently, $(r_+,r_-)$.
To parametrize $\hat{M}_q$ consistent
with the above constraint, consider a pair of 2-vectors,
\begin{equation}
\vec{\hat{q}}_{\phi} =  (\hat{q}^0, \hat{q}^1) = ( f cosh {\phi},
f sinh {\phi})
\nonumber \end{equation}
\begin{equation}
\vec{\hat{q}}_t = ( \hat{q}^2, \hat{q}^3) = \left( \sqrt{f^2 -
l^2} cosh(t/l),
\sqrt{f^2 - l^2} sinh(t/l) \right) 
\end{equation}
where $ f= f(r)$, with $r$ a radial coordinate which for an
appropriate $f(r)$ will
become the radial coordinate appearing in the line element for
the BTZ black hole. As far the
constraint given by Eq. 34 is concerned, the functional form of
$f(r)$ is
irrelevant.
Computing the line element in terms of the parameters
$(t/l,r,\phi)$, we get
\begin{equation}
ds^2 = (\frac{f^2}{l^2} - 1) dt^2 -
\frac{f^{'\;2} dr^2}{(\frac{f^2}{l^2} - 1)} 
- f^2 d\phi^2 
\end{equation}
where ``prime'' indicates differentiation with respect to $r$.

Anticipating the results to be given below, let us 
compare this line element with that for the
BTZ black hole~\cite{rone}:
\begin{equation}
ds^2= [\frac{r^2}{l^2} -M + \frac{J^2}{4r^2}] dt^2 -
\frac{dr^2}{[\frac{r^2}{l^2} -M + \frac{J^2}{4r^2}]} - r^2 [d\phi
- \frac{J^2}{2r^2}dt]^2 
\end{equation}
If we identify the labels $M$ and $J$ with the Casimir invariants
of an irreducible representation of the AdS group as discussed in
the previous sections,
we see that the line element given by Eq. 37 corresponds to an
irreducible representation with $J= 0$ and $M = 1$. Such a
parametrization cannot provide us with an AdS black hole with a
continuous range of values for $J$ and $M$. What we need is a
space-time manifold $M_q$ in which the coordinates $q^A$ carry an
arbitrary irreducible representation of the AdS group. We will
construct this manifold explicitly by performing appropriate
gauge
transformations on $\hat{M}_q$. Although the original theory was
invariant under  $SL(2,R) \times SL(2,R)$ gauge transformations,
we have already
reduced this symmetry by choosing to work in a gauge in which
the left over symmetry is just $SO(1,1) \times SO(1,1)$
generated, respectively, by $J_2^{\pm}$. It turns out to be more
convenient to work with generators
\begin{equation}
J_2 = J^+_2 + J^-_2 ;\;\;\;\;\;\;\;\;\;\; l\Pi_2 = J^+_2 - J^-_2
\end{equation}
As was the case with the compact generators $J_0$ and $\Pi_0$,
the non-compact operators $J_2$ and $\Pi_2$ generate a $SO(1,1)
\times SO(1,1)$ subgroup of the AdS group.
We identifying the
parameters $\phi$ and
$t/l$, respectively, with each $SO(1,1)$ and proceed in the same
manner as we did for the compact generators in
reference~\cite{rseven}. In particular,
consider the local gauge
transformation~\cite{rseven,reight,rtwenty,rtwoone,rtwotwo}
\begin{eqnarray}
\vec{q}_{\phi'}(\phi, t/l) = e^{ \left( \frac{r_+}{l}
\phi - \frac{r_-t}{l^2} \right)J^2} \vec{\hat{q}}_{\phi} ( \phi)
\nonumber \\
\vec{q}_{t'} (t/l,\phi) = e^{ \left( \frac{r_-}{l} \phi
- \frac{r_+t}{l^2} \right)l\Pi^2} \vec{\hat{q}}_{t} (t/l)
\end{eqnarray}
It then follows that
\begin{eqnarray}
\vec{q}_{\phi '}(\phi + 2 \pi, t/l) = e^{ 2
\pi\frac{r_+}{l} J^2} \vec{q}_{ \phi '} ( \phi, t/l)
\;\;\;\;\;\;\;\;\;\;\nonumber \\
\vec{q}_{t '}(t/l,\phi + 2 \pi) = e^{ 2
\pi\frac{r_-}{l} l\Pi^2} \vec{q}_{t} (t/l, \phi)
\;\;\;\;\;\;\;\;\;\;\;\nonumber \\
\vec{q}_{\phi '}(\phi + 2 \pi, t/l + 2 \pi) = e^{ 2 \pi
\left( \frac{r_+}{l}  - \frac{r_-}{l} \right) J^2}
\vec{q}_{\phi'} ( \phi, t/l)\nonumber \\
\vec{q}_{t'}(t/l +2\pi,\phi + 2 \pi) = e^{ 2 \pi
\left(\frac{r_-}{l} - \frac{r _+}{l} \right)l\Pi^2}
\vec{q}_{t'} ( \phi, t/l) \end{eqnarray}
The factors by which these quantities change as $\phi \rightarrow
(\phi + 2\pi)$ are reminiscent of the holonomies which we
obtained in section 4. There we found that these holonomies in
$M$ led to a discrete identification group. To be consistent with
this identification group, we must require that the hyperbolic
variable $\phi$ be periodic. Thus, the periodicity of phi
follows from the topology of $M$ in the presence of a source.
With this provision, the vector $(\vec{q}_{\phi'}, \vec{q}_{t'})$
transforms in the same way as the one which in section 4 was
parallel transported around a loop in the manifold $M$.
Calling the
manifold to which such vectors belong $M_q$, we see that
this
manifold incorporates the same dynamics as the phase space
variables in
$M$, and we are justified in using the same letter $q$ for both.
Thus, we can parametrize the manifold $M_{q}$ as
follows~\cite{rone,rseven,rnine}:
\begin{eqnarray}
q^0 = f cosh \left( \frac{r_+}{l}\phi - \frac{r_-t}{l^2}
\right)\;\;\;\;\;\;\;\;\nonumber \\
q^1 = f sinh \left( \frac{r_+}{l}\phi - \frac{r_-t}{l^2}
\right)\;\;\;\;\;\;\;\;\nonumber \\
q^2 = \sqrt{ f^2 - l^2 } cosh \left( \frac{r_-}{l}\phi -
\frac{r_+t}{l^2} \right)\nonumber \\
q^3 = \sqrt{ f^2 - l^2 } sinh  \left( \frac{r_-}{l}\phi -
\frac{r_+t}{l^2} \right) \end{eqnarray}
From these we can compute the line element. It is easy to show
that it is the same as
that given by Eq. 38 for $r > r_-$.

\section{Supersymmetric Black Holes}
The theory which we will describe below is the supersymmetric
generalization of the theory which was discussed in the previous
sections. We will see that the emerging 
macroscopic theory consists of a supermultiplet of ordinary
space-times and, as a special case of this, a supermultiplet
consisting
of black holes only.
The simplest way of obtaining a supersymmetric extension of the
anti-de Sitter group is
to begin with the AdS group in its $SL(2,R) \times SL(2,R)$
basis. The $N=1$ supersymmetric extension of each $SL(2,R)$
factor is
the supergroup $OSp(1|2;R)$. Thus, one arrives at the (1,1) form
of the $N=2$ super AdS group. Its algebra is given by
\begin{eqnarray}
[J_a^{\pm}, J_b^{\pm}] = -i\epsilon_{ab}^{\;\;\;c} J_c^{\pm};
\;\;\; 
[J^{\pm}_a,Q^{\pm}_{\alpha}] =
-\sigma^{a\;\beta}_{\alpha}Q^{\pm}_{\alpha}; \;\;\; 
\{Q^{\pm}_{\alpha}, Q^{\pm}_{\beta} \} =
-\sigma^{a}_{\alpha\beta}
J^{\pm}_a \;\;\;\;\;\nonumber \\ 
\{Q^+_{\alpha}, Q^-_{\beta}\} = 0;  \;\;\;\;\; [J^+,J^-]= 0
\;\;\;\;\;\;\;\;\;\;\;\;\;\;\;\;\;\;\;\;\;\;\;\;\;\end{eqnarray}
The Casimir invariants are given by
\begin{equation}
C_{\pm} = j_{\pm}^2 +
\epsilon^{\alpha\beta} Q^{\pm}_{\alpha} Q^{\pm}_{\beta}
\end{equation}
The spinor indices are raised and lowered by antisymmetric metric
$\epsilon^{\alpha \beta}$ defined by
$\epsilon^{12} = -\epsilon_{12} = 1 $. The matrices
$(\sigma^a)_{\alpha}^{\beta}$, $( a= 0,1,2)$, form a
representation of 
$SL(2,R)$ and satisfy the Clifford algebra
\begin{equation}
\{\sigma^a,\sigma^b\} = \frac{1}{2}\eta^{ab}\end{equation}
More explicitly, we can take them to be of the form:
\begin{eqnarray}
\sigma^0 = \frac{1}{2} \left(\begin{array}{cc} 1 & 0 \\ 0 & -1
\end{array} \right) ;\;\;\;\; 
\sigma^1 = \frac{1}{2} \left(\begin{array}{cc}  0 & i \\ i & 0
\end{array} \right) ;\;\;\;\; 
\sigma^2 = \frac{1}{2}\left(\begin{array}{cc} 0 & 1 \\ -1 & 0
\end{array} \right)
\end{eqnarray}
It is important to note that the supersymmetry generators of
$OSp(1|2,R)$ do not commute with the Casimir invariant of its
$SL(2,R)$ subgroup. That is,
\begin{equation}
[j^2_{\pm}, Q_{\alpha}] \neq 0
\end{equation}
 
Since super AdS group is semi-simple, we can construct its
irreducible representations by first constructing the irreducible
representations of $OSp(1|2,R)$~\cite{rthirteen}. Depending on
which $OSp(1|2,R)$
we are considering, the states within any such supermultiplet are
the corresponding irreducible representations of 
$SL(2,R)$ Characterized by the Casimir invariants $j_+$ or $j_-$,
respectively. Based on the rationale given for the non-
supersymmetric case~\cite{rseven}, the irreducible
representations of interest
for the present case are those which can be obtained from the
unitary
discrete series of $SL(2,R)$ and which are bounded from below. To
construct the supermultiplet corresponding to,
say, the ``plus'' generators in Eq. 22, we can take the Clifford
vacuum state
$|\Omega^+>$ to be the $SL(2,R)$ state with the lowest eigenvalue
of $J_2^+$. In the notation of Eq. 22, this corresponds to an
$m=0$ state:
$$|F_+, m > = |F_+, m=0 > = |F_+ >$$
Then, the superpartner of this state, again with $m=0$, is the
state $|F_+ +1/2>$ obtained by the application of one of the
$Q$'s. The corresponding values of $j^2_+$ are $F_+(F_+ -1)$ and
$(F_+ +1/2)(F_+ -1/2)$, respectively. The supermultiplet for the
second $OSp(1|2,R)$ can be constructed in a similar way.

We are now in a position to construct the (1,1) super AdS
supermultiplet as a direct product of the two $OSp(1|2,R)$
doublets. Altogether, there will be four states in the
supermultiplet. They will have the following labels:
\begin{equation}
|F_+,F_- >; \hspace{0.4cm} |F_+ + 1/2,F_- >;
\hspace{0.4cm} |F_+, F_- + 1/2>; \hspace{0.4cm}
|F_+ + 1/2, F_- + 1/2 > \end{equation}
From these, we can also obtain the  expressions for the
eigenvalues $(M,J)$ of various states within the supermultiplet:
\begin{eqnarray}
|M_1, J_1 > = |M, J > \nonumber \\
|M_2, J_2 > = |M + 2F_+ - 1/2, J + 2F_+ - 1/2 > \nonumber \\
|M_3, J_3 > = |M + 2F_- - 1/2, J - 2F_- + 1/2 > \nonumber \\
|M_4, J_4 > = |M + 2(F_+ + F_-) - 1, J + 2(F_+ - F_-) >
\end{eqnarray}
these states transform into one another under supersymmetry
transformations.

The Chern Simons action for simple and semisimple supergroups has
the same structure as that for Lie groups. The only difference is
that the trace operation is replaced by super trace (Str)
operation. So, in the $OSp(1|2,R)
\times OSp(1|2,R)$ basis the
Chern Simons action for the super AdS group has the same form as
that given by Eq. 14. But now the
expression for connection is given by
\begin{equation}
A^{\pm} = \left[ A^{\pm\;a}_{\mu} J^{\pm}_a +
\chi^{\pm\alpha}_{\mu}
Q^{\pm}_{\alpha}\right]dx^{\mu}\end{equation}
Just as in the non-supersymmetric case, to have a
non-trivial theory, we must couple sources to the Chern Simons
action. To do this in a gauge
invariant and locally supersymmetric fashion, we must take a
source to be an irreducible representation of the super AdS
group. As we saw above, such a supermultiplet
consists of four AdS states. To couple it to the gauge fields, we
must first generalize the canonical variables we used in the AdS
theory to their supersymmetric forms~\cite{rthirteen}:
\begin{equation}
 \Pi_A \rightarrow (\Pi_A,\Pi_{\alpha}) \hspace{0.5in} q_A
\rightarrow (q_A,q_{\alpha})\end{equation}
Then, the source coupling can be written as
\begin{equation}
I_s = \int_C \left[\Pi_Adq^A + \Pi_{\alpha}dq^{\alpha} + (A^+ 
+ A^-)+ constraints \right] \end{equation}
where again $C$ is a path in $M$.
The constraints here include those discussed for the AdS group
and, in addition, those which relate the AdS labels of
the Clifford vacuum to the Casimir eigenvalues of the super AdS
group. The combined action
\begin{equation}
I = I_{cs} + I_s \end{equation}
leads to the constraint equations
\begin{equation}
\epsilon^{ij} F^{\pm\;a}_{ij} = 2\pi a_{\pm} J^{\pm\;} \delta^2
(\vec{x},\vec{x^0}); \hspace{0.5in} 
\epsilon^{ij} F^{\pm\alpha}_{ij} = 2\pi a_{\pm} 
Q^{\pm\alpha}\delta^2(\vec{x},\vec{x^0}) \end{equation}

Up to this point, everything proceeds in parallel with the non-
supersymmetric case. Differences begin to show up when one
attempts to solve these equations by
choosing a gauge again so that the gauge symmetry is reduced to
$SO(1,1) \times SO(1,1)$:
\begin{equation}
\epsilon^{ij}F^{\pm\;2} = 2\pi a_{\pm} J_2^{\pm}
\delta^2(\vec{x},\vec{x^0}) \end{equation}
Although this equation is identical in form to Eq. 27 for the
non-supersymmetric case, there is an essential difference in the
underlying physics. In the supersymmetric case, the
supermultiplet which we couple to the Chern Simons action consist
of four $SO(2,2)$ states with different values of $F_{\pm}$. As a
result, in the parallel transport of $q^A$ around a close path
analogous to the non-supersymmetric case, there will be four sets
of holonomies with different values of $(j_+, j_-)$ or,
equivalently, $(r_+, r_-)$. Moreover, in the non-supersymmetric
case, a single source with Casimir invariants $(r_+, r_-)$ or,
equivalently, $(M,J)$ will give rise to an AdS black
hole~\cite{rseven} for which the 
line element is characterized with the corresponding values of
$M$ and $J$ and has the form given by Eq. ?.
In the supersymmetric case, the source is a supermultiplet in
which there are four states of differing  $(M,J)$ values.
Then, 
depending on which set $(M,J)$ that we choose, we will get a
different BTZ solution. Since $M$ and $J$ are not invariant under
supersymmetry transformations, these solutions are transformed
into each other under supersymmetry.
This makes it
impossible for a single c-number line element of the type given
by Eq. 38 to
correspond to all the AdS states of a supermultiplet.

The situation
here runs parallel to what was encountered in connection with
super Poincar\'e Chern Simons theory~\cite{rtwenty}. There it
was
pointed out that standard classical geometries were not capable
describing these
structures and that one must make use of {\it nonclassical
geometries}. Such
geometries can be described in terms of three elements: 

{\bf 1}. An algebra such as a Lie algebra or a Lie superalgebra. 

{\bf 2}. A line element operator with values in this algebra. 

{\bf 3}. A Hilbert space on which the algebra acts linearly. 

For the problem at hand, the algebra of interest is the $N =
(1,1)$ super AdS algebra in $2+1$ dimensions. The corresponding
Hilbert space is the representation space of the superalgebra
given by Eq. 48. Then, instead of the BTZ line element given,
 we begin with a line element operator with values in the
$N = (1,1)$ superalgebra and assume that its diagonal elements
depend on the algebra only through the Casimir operators
$(\hat{M},\hat{J})$ of its $SO(2,2)$ subalgebra. The ``hats'' on
top of $M$ and $J$ are meant to distinguish the operators from
the corresponding eigenvalues. Thus, we have
$$ds^2 = ds^2(\hat{M}, \hat{J})$$
The matrix element of this operator for each state of the
supermultiplet will produce a c-number line element:
\begin{equation}
<M_k, J_k| ds^2 (\hat{M}, \hat{J}) |M_k, J_k> = ds^2(M_k,J_k)
\end{equation}
In other words, for each state of the supermultiplet, the
nonclassical geometry produces a layer of classical space-time.
The number of the layers is equal to the dimension of the
supermultiplet. Supersymmetry transformations act as messengers
linking different layers of this multilayered space-time

An equivalent way of constructing the operator line element is to
begin with the parametrization of $q^A$ given by Eq. 41. Then,
replacing the Casimir eigenvalues in those expressions with the
corresponding Casimir operators, we can proceed to compute the
line element operator. The result will look like the line element
given by Eq. 38, except that now Casimir eigenvalues are replaced
with Casimir operators. This means that, for
consistency, we must also interpret the quantities $J_2^{\pm}$ in
Eq. 54 as operators. Acting on different states of the
supermultiplet, they will give the corresponding $F_{\pm}$
eigenvalues. There will therefore be not one set but four sets of
holonomies $W[A^+]$ and $W[A^-]$. Each set will produce the
discrete identification subgroup in the corresponding layer of
space-time. 

Consider, next, the conditions under which every layer of the
supermultiplet corresponds to an AdS black hole. For this to be
true, we must have
$$M_k \geq 0;\;\;\;\;\; |J_k| \leq lM_k$$
This, in turn implies that
$$F_+ \geq F_- \geq 1$$
In the notation of Eq. 48, for $|J| = lM$, two layers of the
supermultiplet become extreme AdS black holes. The only exception
is in the limiting case when $M=J=0$, in which case there will be
three extremal black holes in the supermultiplet. It is also
interesting to note that for an appropriate choice of $M$ and $J$
or, equivalently, $F_+$ and $F_-$, the same supersource which
generates a black hole in one layer can generate a naked
singularity in another. 

\bigskip
This work was supported, in part by the Department of Energy
under the contract number DOE-FGO2-84ER40153. We would like to
thank the Organizing Committee of Johns Hopkins Workshop for
their
hospitality and for the opportunity to present this material. 
We have also benefited from the hospitality of
Aspen Center for Physics, where part of this work was carried
out.


\vspace{0.1in}

\begin{thebibliography}{99}
\bibitem{rone} M. Ba\~{n}ados, C. Teitelboim and J. Zanelli,
 Phys. Rev. Lett. {\bf 69} (1992) 1849; M. Ba\~{n}ados, M.
Henneaux, C. Teitelboim and J.
Zanelli, Phys. Rev. D {\bf 48} (1993) 1506.
\bibitem{rtwo} A. Achucarro, P. Townsend, Phys. Lett.B
{\bf180} (1986) 35
\bibitem{rthree} E. Witten, Nucl. Phys {\bf B311} (1988) 46; {\bf
B323} (1989) 113
\bibitem{rfour} S. Carlip, Phys. Rev. {\bf D51} (1995) 632
\bibitem{rfive} A.P. Balachandran, L. Chandar, A. Momen, Nuc.
Phys. {\bf B461} (1996) 581; eprint archive gr-qc/9506006
\bibitem{rsix} A. Strominger
\bibitem{rseven} S. Fernando, F. Mansouri, eprint archive hep-
th/9804147, Int. Jour. Mod. Phys. {\bf A}, {\it in press}
\bibitem{reight} F. Mansouri, M.K. Falbo-Kenkel, Mod. Phys Lett.
{\bf A8} (1993) 2503; F. Mansouri, Proceedings of XIIth Johns
Hopkins Workshop, ed. Z. Horwath, World Scientific, 1994
\bibitem{rnine} S. Carlip, eprint archive gr-qc/9506079; Clas.
Quan.
Grav. {\bf 12} (1995) 2853
\bibitem{rten} P.C. Aichelburg, R. Guven, Phys. Rev. {\bf D 24}
(1981) 2066
\bibitem{releven} R. Kallosh, A. Linde, T. Ortin, A. Peet, A. Van
Proyen Phys. Rev. {\bf D 46} (1992) 5278
\bibitem{rtwelve} O. Coussaert, M. Henneaux, eprint archive hep-
th/9310194; Phys. Rev. Lett. {\bf 72} (1993) 183
\bibitem{rthirteen} J.M. Izquierdo and P.K. Townsend, eprint
archive gr-qc/9501018; Clas. Quan. Grav. {\bf 12} (1995) 895
\bibitem{rfourteen} R. Kallosh, D. Kaster, T. Ortin, T. Torma,
Phys.
Rev. {\bf D 50} (1994) 6374
\bibitem{rfifteen} M.J. Duff, J. Rahmfeld, Phys. Lett. {\bf B
345}
(1995) 441; eprint archive hep-th/9605085; M.J. Duff, H. Lu, C.N.
Pope, Phys. Lett. {\bf B
409} (1997) 136
\bibitem{rsixteen} A.R. Steif, Phys. Rev. Lett. {\bf 69} (1995)
1849
\bibitem{rseventeen} R. Kallosh, eprint archive hep-th/9503029;
Phys.
Rev. {\bf D 52} (1995) 1234
\bibitem{reighteen} T. Ortin, eprint archive hep-th/9705095
\bibitem{rnineteen} S. Fernando, F. Mansouri, eprint archive hep-
th/9809139
\bibitem{rtwenty} Sunme Kim, F. Mansouri, Phys. Lett. {\bf
B397}
(1997) 81; F. Ardalan, S. Kim, F. Mansouri, Int. Jour. Mod.
Phys., {\bf A12} (1997) 1183
\bibitem{rtwoone} K. Koehler {\it et al}, Nucl. Phys. {\bf B348}
(1990) 373
\bibitem{rtwotwo} C. Vaz and L. Witten, Phys. Lett. {\bf B327}
(1994) 29

\end{thebibliography}
\end{document}